\documentclass[prb,twocolumn,amsmath,amssymb]{revtex4}
\begin{document}
\author{Kevin Leung,$^{1*}$ Susan B.~Rempe,$^1$ Peter~A.~Schultz,$^1$
Eduardo~M. Sproviero,$^2$ Victor S.~Batista,$^2$ Michael E.~Chandross,$^1$
and Craig~J.~Medforth$^1$}
\affiliation{$^1$Sandia National Laboratories, MS 1415, 1110, 0310, 1411,
\& 1349, Albuquerque, NM 87185\protect\\
$^2$Department of Chemistry, Yale University, P.O.~Box 208107, New Haven,
CT 06520-8107 \protect\\
$^*$Email: {\tt kleung@sandia.gov}
}
\title{Density functional theory and DFT+U study of transition metal
porphines adsorbed on Au(111) surfaces and effects of applied electric fields}

\input epsf

\begin{abstract}
 
We apply Density Functional Theory (DFT) and the DFT+U technique to study
the adsorption of transition metal porphine molecules on atomistically
flat Au(111) surfaces.  DFT calculations using the Perdew-Burke-Ernzerhof
(PBE) exchange correlation functional correctly predict the palladium
porphine (PdP) low-spin ground state.  PdP is found to adsorb
preferentially on gold in a flat geometry, not in an edgewise geometry,
in qualitative agreement with experiments on substituted porphyrins.
It exhibits no covalent bonding to Au(111), and the binding energy
is a small fraction of an eV.  The DFT+U technique, parameterized to B3LYP
predicted spin state ordering of the Mn $d$-electrons,
is found to be crucial for reproducing the correct magnetic moment
and geometry of the isolated manganese porphine (MnP) molecule.  Adsorption
of Mn(II)P on Au(111) substantially alters the Mn ion spin state.
Its interaction with the gold substrate is stronger and more site-specific
than PdP.  The binding can be partially reversed by applying an electric
potential, which leads to significant changes in the electronic and
magnetic properties of adsorbed MnP, and $\sim 0.1$~\AA\, changes in
the Mn-nitrogen distances within the porphine macrocycle.  We conjecture
that this DFT+U approach may be a useful general method for modeling
first row transition metal ion complexes in a condensed-matter setting.
 
\end{abstract}
 
\maketitle
 
\renewcommand{\thetable}{\arabic{table}}
\section{Introduction}

Metalloporphyrins (of which metal
porphines are the simplest examples) are stable molecules which
exhibit a wide range of optoelectronic, magnetic, and mechanical
properties.\cite{porph_review}  Deposited and/or self-assembled
on metal electrodes, porphyrins are attractive candidates for novel
molecular sensors,\cite{sensor} memory,\cite{memory}
and light-harvesting components.\cite{light1,light2}
Transition metal porphyrins are particularly interesting because of
the multiple spin/electronic states available to them.  For example,
ligating the metal ions, or simply depositing the molecules on electrodes,
can preferentially stabilize one state versus another.  Understanding
the detailed molecular interactions responsible for binding
metalloporphyrins to well-characterized metal surfaces is therefore a
subject of great technological interest, one which has received extensive
experimental\cite{sensor,memory,light1,light2,mol_elec1,nature,sol1,sol2,mol_elec2,mol_elec3,mol_elec4}
and theoretical\cite{parrin_sur,par_clone} study.

Memory and sensor applications of transition metal porphyrins rely on
electrochemically-induced switching of the electronic and magnetic
states.\cite{sensor,memory}  The field dependent self-assembly and
mobility of porphyrin molecules on electrodes have also been the subject of 
experimental interest under ultra-high vacuum conditions\cite{mol_elec1,nature}
and in solution.\cite{sol1,sol2}
Electric-field induced changes in the electronic
states can lead to structural changes in porphyrin
molecules, and this may impact their functions and reactivity
in subtle ways.\cite{shaik}
Macrocycle conformational changes in nickel and zinc
porphyrins have recently been the subject of a series of
investigations.\cite{shel1,shel2,shel3}
Attaching ligands to Ni(II) or photoexcitation can
switch the low-spin nickel ground state with a $(d_{z^2})^2$ configuration
to the high-spin expanded $(d_{z^2})(d_{x^2-y^2})$ state,
so that core expansion causes the 
porphyrin ring to undergo a conformational change
from a ruffled to a dome-like geometry.
When outlying porphyrin appendages or bridles are attached to
the Ni porphyrins, this small but energetic motion
triggered by nickel spin state changes can potentially be harvested
as nanomechanical motion in molecular switches.\cite{shel1,shel2}
It would be of great interest to extend this general principle to other
transition metal porphyrins and related molecules deposited on gold
electrodes.  In such cases, the transition metal charge or spin states
can be altered by electrochemical means, and the resulting conformational
changes can be monitored using atomic force microscopy.

Given the wide range of interest in porphyrins adsorbed on metal surfaces,
it is important to elucidate the
precise nature of metalloporphyrin adsorption on and interaction with
metal electrodes.  Experimentally, it is known that certain substituted
porphyrins and related molecules lie flat on Au(111)
surfaces.\cite{mol_elec1,nature,sol1,sol2,mol_elec2,mol_elec3}
Even when they contain bulky substitutents, the
molecules are only slightly distorted from a planar
adsorption geometry.\cite{mol_elec4}  It is also possible to
have substituted porphyrins self-assembled into a tube-like geometry
and lie edgewise on material surfaces if they
are tethered using sulfide or other linkages.\cite{light1,sol1}

Density functional theory (DFT) might appear to be the theoretical
method of choice to shed light on the binding energies,
spin states, geometries, and external electric field effects
of porphyrins adsorbed on metal surfaces.
DFT simultaneously addresses the electronic and geometric properties
of the composite molecule-metal system.  It is a formally exact method,
but the quality of its predictions depends on the approximate exchange
correlation functional used.  An early DFT
calculation\cite{parrin_sur} was performed prior to detailed
experimental studies of porphyrin adsorption on atomistically flat
gold surfaces.\cite{mol_elec1,mol_elec2,mol_elec3,mol_elec4}
(For the purpose of this work, we ignore the
herring-bone surface reconstruction on Au(111),\cite{au_expt} which occurs
on long length-scales.)
Motivated by molecular electronics experiments on self-assembled
molecules which are somewhat similar to porphyrins,\cite{stack_expt}
Lamoen {\it et al}.~\cite{parrin_sur} examined palladium porphines
(PdP) adsorbed edgewise on
Au(111), and reported a large, $\sim 10$~eV, binding energy.  This calculation
was performed using the local density approximation (LDA), which tends
to overestimate the binding energy.  A subsequent DFT work\cite{par_clone}
considered PdP adsorbed in a similar edgewise geometry on Al(111)
surfaces.  This study reported a sub eV adsorption energy using
LDA, and an even smaller binding
energy using the Perdew-Burke-Ernzerhof (PBE)\cite{pbe} version of the
generalized gradient approximation (GGA) to the exchange correlation
functional.  GGA is generally more accurate than
LDA for treating molecule-material binding\cite{moroni} and
has been widely used in such studies.\cite{selloni}

Guided by recent experiments and building on these earlier DFT results,
one of the goals of this work is to apply the PBE functional
to re-examine PdP adsorption on the Au(111) surface.  
We consider both the flat adsorption geometry found in
experiments where porphyrins are not tethered to metal
surfaces,\cite{mol_elec1,mol_elec2,mol_elec3,mol_elec4} and
the edgewise geometry adopted in earlier calculations.

PdP can be considered a prototypical metalloporphyrin which adsorbs weakly on
Au(111).  For many transition metal compounds and materials, such as
second row elements like Pd, most implementations of DFT appear to yield
the correct ground spin state and electronic properties, and they are
valuable techniques to apply.\cite{second_row}
However, we are also interested in first row transition
metal porphyrins (e.g., Ni(II) and Mn(II), which are pertinent for studying
effective ion size-driven conformational changes in porphyrin
molecules).\cite{shel1,shel2,shel3}  Here we encounter a dilemma that has
hindered DFT studies of composite systems involving first row transition
metal ions.  In brief, DFT implementations with non-hybrid exchange correlation
functionals such as PBE\cite{pbe} and PW91,\cite{pw91} widely used in the
condensed matter physics community,
treat gold and other metal surfaces accurately but are less accurate for
interactions between ligand/crystal field and first row transition
metal ions.  The Becke 3-parameter Lee-Yang-Parr (B3LYP)\cite{b3lyp}
hybrid exchange correlation functional correctly
predicts the high-spin ground state for MnP, but is at present
too costly for modeling metal surfaces.\cite{cost,metal}

Accurate predictions for bulk transition metals lattice constants and
magnetic properties were among the early successes of non-hybrid
GGA.\cite{gga} PW91 and PBE are among this class of GGA.
However, these functionals underestimate the exchange interaction among
the strongly localized, partially filled $3d$ orbitals in first row
transition metal {\it ions}.  This leads to underestimation of the
stability of high-spin states in some first row transition
metal ions in impurity centers or ligand fields, where non-hybrid GGA
often predicts ground states with incorrect spin
multiplicities.\cite{reiher1,reiher2,leung01}
Indeed, a recent {\it ab initio} molecular dynamics simulation of Mn(II)
centers was compelled to apply a high-spin constraint throughout the
trajectory.\cite{klein}  

Hybrid exchange correlation functionals contain fractional 
non-local or Hartree-Fock exchange, which amounts
to 20\% for B3LYP.\cite{b3lyp}  This 
admixture is apparently the right amount
to reproduce the experimental spin ordering and energy splitting
between spin states in many complexes between
ligands and Mn\cite{ed1,ed2,ed3} as well as other first row transition metal
ions,\cite{ed3} and is widely and
successfully used for Mn centers.\cite{ed1,ed2,ed3,isobe,devisser,khav}
A slightly smaller admixture yields better results in other
cases.\cite{reiher1,reiher2} At present, applying these hybrid
functionals in simulation cells with periodic boundary condition
is computationally costly---up to 100 times more so than for
non-hybrid functionals.\cite{cost} This factor may further depend
on system size.  As such, B3LYP has only seen preliminary applications
in condensed phase systems.\cite{prelim}  To our knowledge, it has
not been applied to the large, slab-like, periodically replicated
supercells needed to study porphyrin adsorption on metal
surfaces,\cite{metal} or for that matter, for large scale
DFT calculations that require periodic boundary
conditions, such as in aqueous systems via {\it ab initio}
molecular dynamics.\cite{klein} As will be shown, periodically replicated
simulation cells with adequate Brillouin zone sampling are crucial for
modeling porphyrin adsorption on gold surfaces.

We note that hybrid functionals are not universally successful
for all transition metal species.  For example, (1) hybrid functionals
tend to overestimate the stability of $Ns^1(N-1)d^{n+1}$ electronic
configurations over $Ns^2(N-1)d^n$ ones in transition metal
atoms.\cite{cpl}  (So does non-hybrid GGA.\cite{truhlar})  This
consideration does not play a role when the metal $s$-orbital is
no longer available due to its involvement in doubly occupied
bonding orbitals,\cite{devisser} or in Mn(II)P and Mn(III)P, which
have largely empty Mn $4s$ orbitals.  (2) The binding energies
and bond lengths of many transition metal dimers are more accurately
predicted using non-hybrid rather than hybrid GGA.\cite{truhlar}
This issue may be related to (1).  Thus, complexes with Mn-Mn bonds may
be particularly challenging for hybrid functionals.\cite{ed4,spro}
(3) For as yet unknown reasons, B3LYP is ambiguous regarding the stability
of the high-spin electronic configurations of some first row transition metal
porphines, such as ligated FeP,\cite{ghosh_rev,ghosh4} while non-hybrid
GGA performs considerably worse;\cite{ghosh_rev} a larger admixture
of Hartree-Fock exchange seems necessary in this case.

A promising alternative method to characterize the correct spin state
is to use Quantum Monte Carlo (QMC) methods, which up to now
have been used to study porphine (H$_2$P) which lacks coordinated
metal ions.\cite{qmc} Another option is to impose the experimentally known
spin polarization in the entire supercell.\cite{klein}  As will be
discussed, however, this approach may not yield accurate metal-nitrogen
distances within the porphyrin ring.  Furthermore,
when a metal electrode is present, the bulk metal valence electrons
can be excited to the conduction band to give arbitary 
spin multiplicities at negligible energy penalties.
So, a global spin constraint will likely change the bulk
metal magnetic behavior without affecting the adsorbed transition metal
ion complex.  Clearly, new computational techniques are needed
to treat porphyrin molecules adsorbed on metals.

In this work, we apply the DFT+U method\cite{ldau0,ldau1} to treat the
composite manganese porphine-gold metal system.  This method, which
emphasizes the role of on-site screened coulomb interactions, has
successfully predicted the correct electronic ground state for transition
metal oxide crystals such as NiO\cite{nio} and LaCoO$_3$,\cite{lacoo3}
where traditional non-hybrid GGA has failed.
This mean field approach augments the DFT exchange correlation interactons
{\it among electrons in a subset of orbitals} (Mn $3d$ in our case) with
Hartree-Fock-like interactions parameterized with coulomb
($U$) and exchange ($J$) terms.  Electrons in these orbitals still
interact with the rest of the system (including the C, H, N atoms in
the porphine ring and the Au(111) substrate) 
via the DFT formalism.  The specific implementation
we use is described in more detail in the Supporting Information section.

Formally speaking, the parameters $U$ and $J$ are related to coulomb and
exchange integrals.\cite{nio,pickett}  In practice,
$U$ has been set at values required to achieve agreement with
experiments.\cite{rohrbach} 
When dealing with molecules embedded with transition metal ions,
we propose that $U$ and $J$ can be fitted to either experiments or
predictions via other gas phase theoretical methods such as high level
quantum chemistry, quantum Monte Carlo, and hybrid functional
DFT---as long as the latter are known to agree with
experiments.  The DFT+U technique is readily amenable to periodically
replicated simulation cells, and is thus well suited
for studying metal porphine molecules adsorbed on gold surfaces.  We note
that this method is empirical in nature, and it assumes that correlation
between the $3d$ electron can be approximated as static.\cite{singh}
Nevertheless, it has seen wide and successful applications to
many solid state materials where LDA/GGA treatments
fail for electronic/magnetic properties.\cite{ldau0,nio,lacoo3,rohrbach}
It may also become a general technique for {\it ab initio}
molecular dynamics study of first row transition metal centers
in condensed phases.\cite{klein,water}

As a proof of principle, we will focus on a second prototype porphyrin:
manganese (II) porphine (MnP).  MnP exhibits behavior drastically different
from that of PdP discussed above.  PdP is umambigously stable in the low spin
state, and PBE and DFT+U methods yield similar predictions.  However,
Mn(II) complexes in general and Mn(II) porphyrins in particular are known
to have high-spin ground states.\cite{mn_expt,mn_expt1,siegbahn}
We will show that PBE incorrectly predicts the intermediate spin state
to be the MnP ground state.  On the other hand, by fitting $U$
to B3LYP results (which yield the experimental high spin ground state),
both the high/intermediate spin energy splitting and the gas phase
MnP geometry are successfully reproduced using the DFT+U method.\cite{spectra}

Having validated the DFT+U technique for isolated MnP, we further
apply it to investigate the geometric, electronic, and magnetic properties
of MnP adsorbed on Au(111) surfaces.  We also examine the effect of an
electric field on the adsorbed MnP, and show that it can induce
significant changes in these properties.
The MnP structure is particularly strongly affected by the applied
field, with the Mn-N distance in the porphine ring increasing by
0.13~\AA.  This suggests that Mn porphyrins may exhibit 
conformational changes related to those seen for Ni
porphyrins.\cite{shel1}

To summarize, we have conducted a comprehensive study of the adsorption of
two prototype transition metal porphine molecules, PdP and MnP, on gold
surfaces.  The effect of an applied electric field on the
properties of adsorbed MnP is investigated.
We also emphasize the success and importance of applying the DFT+U technique
to treat the first row transition metal ion Mn(II), and conjecture
that this technique may be widely applicable to condensed phase systems.

This paper is organized as follows.  Section~II describes the methods
and models used in our calculations.  Section~III details the
results on isolated porphine molecules and porphines adsorbed on
Au(111).  Section~IV concludes the paper with further
discussion of the results obtained.
 
\section{Methods and models}
 
DFT/PBE and DFT+U calculations are performed using the Vienna {\it ab initio}
Simulation Package (VASP),\cite{vasp} the projector augmented
waves method,\cite{blochl} and associated pseudopotentials.\cite{paw}
For details of the DFT+U implementation, see Bengone {\it et al.}\cite{ldau2}
and the Supporting Information.  The cutoff for wave functions is set
to 400~eV.  LDA calculations are performed
with both VASP and SeqQuest,\cite{seq} using ultrasoft
pseudopotentials\cite{ultra} and norm conserving pseudopotentials,
respectively.  An energy convergence criterion of 
10$^{-3}$~eV for each atomic configuration is enforced.

\begin{figure}
\centerline{ \hbox{\epsfxsize=2.70in \epsfbox{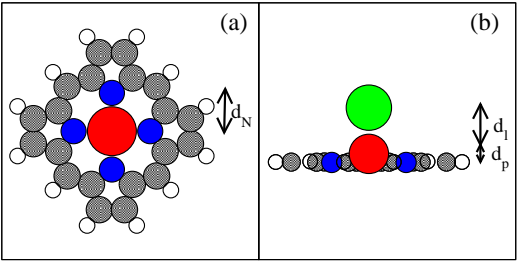}}}
\caption[]
{\label{fig1} \noindent
A metal-porphine molecule, (a) without ligands, (b) attached to a
ligand (Cl$^-$ in this side view example).  Carbon, nitrogen, hydrogen,
metal, and chlorine atoms are depicted as grey, blue, white, red, and
green circles, respectively.  The metal-nitrogen ($d_{\rm N}$),
metal-porphine plane ($d_{\rm p}$), and metal-ligand ($d_{\rm l}$)
distances are illustrated in the figure.  As depicted, the metal
ion in panel (b) is displaced towards the ligand atom, and
$d_{\rm p}$ is defined as positive (``above the macrocycle ring'').
The Au(111) substrate, if present, is below the macrocycle on the
oposite side to the ligand.
}
\end{figure}

Calculations on isolated porphyrin molecules (Fig.~\ref{fig1})
are performed using 16$\times$16$\times$10 \AA$^3$ simulation cells.
For porphines with metal ions
ligated with H$_2$O or Cl$^-$ groups, cell sizes up to
16$\times$16$\times$15 \AA$^3$ are used, and dipole corrections are
applied to remove the coupling between periodic images.\cite{makov,bengt,schul}
These simulation cells converge the spin splittings to 10~meV or better.
 
To model porphines deposited flat on Au(111) surfaces, we use
supercells of lateral size 14.79$\times $15.37~\AA$^2$.
They contain three to six gold layers of 30 Au atoms each,
and we apply various Brillouin sampling schemes.  The cell
lengths in the z-direction are 20, 22, 25, and 28~\AA\, when 
3, 4, 5, and 6 layers of gold Au are present, respectively.
The Au atoms in the bottom layer are fixed at their bulk face center
cubic (FCC) positions.\cite{fcc,au_gga}  Dipole corrections are also applied
for these slab geometry calculations.\cite{makov,bengt,schul}
 
To study edgewise adsorption of PdP on Au(111) using LDA, we adopt the
supercell of
Ref.~13,
except that the cell dimension in
the $z$-direction is increased to 27~\AA\, to minimize the coupling between
periodically replicated images.  The supercell contains three gold layers
totaling 36 Au atoms, and a PdP molecule with two edge protons straddling
a gold atom (see Fig.~\ref{fig2}b).  The equilibrium PBE lattice
constant is larger than that of LDA, and PBE investigations of
this adsorption geometry employ a simulation cell of size
5.91$\times$15.36$\times$27 \AA$^3$.  Monkhorst-Pack grids\cite{mp} of
density up to 4$\times$2$\times$1 are used to sample the Brillouin zone.

Optimizations and single-point energy calculations of gas
phase porphine molecules are computed with the program
Jaguar\cite{jaguar} using the DFT B3LYP hybrid functional\cite{b3lyp}
with the 6-311G** basis set for second row atoms, and
the lacv3p** basis set for Mn that includes Effective Core Potentials
(ECPs).\cite{ecp}
Additional optimization calculations are performed using all-electron
basis sets with the Gaussian~03 code.\cite{g03}
Optimized energies are found with Ahlrichs and co-workers'
all-electron TZV basis\cite{ahlrichs} on the metal ion
and the 6-311G** basis for all other atoms, followed by
a single point energy calculation using
the Wachters-Hay\cite{wachters}
all electron basis set augmented with a diffuse $d$-function, 6-311+G, on
the metal ion, and 6-31G* on all other atoms.

Molecular mechanics calculations are performed with
the LAMMPS molecular dynamics software\cite{lammps} 
modified to utilize the porphyrin potential of Shelnutt and
coworkers.\cite{shel_para} The gold substrate is 6 layers thick,
and has dimensions 41~\AA\, in the $x$ and $y$ directions.
Au atoms are held fixed throughout the simulations; the gold-porphine
interactions are calculated
using previously published force fields.\cite{rappe}
Molecular mechanics total energy minimization calculations are
converged to sub-meV levels.

\section{Results}
 
\subsection{Isolated porphyrin molecules}
 
This section describes the geometries of isolated
porphine molecules, and the energy splitting as a function
of the spin multiplicity, as predicted by the
different exchange correlation functionals.
All symmetries are turned off in PBE and
DFT+U calculations, which are performed using
the VASP code and are converged to a few meV.
B3LYP calculations for MnP enforce D$_{\rm 2h}$
or D$_{\rm 4h}$ symmetries, depending on the Jahn-Teller
distortion dictated by the degeneracy.
Upon binding to a ligand or in the presence of a substantial electric
field, the metal ion displaces out of the porphine ring towards the
ligand atom by a distance $d_{\rm p}$, where the porphine ring position is
taken to be the average $z$ coordinate of the 36 porphine atoms
excluding the metal ion.  The distance between the transition
metal ion and the ligand atom closest to it is defined as $d_{\rm l}$.
These definitions are illustrated in Fig.~\ref{fig1}.
 
\begin{table}\centering
\begin{tabular}{c c c c c c c} \hline
{\bf porphyrin} & method & $E_{\rm rel}$ & $d_{\rm N}$ & 
$d_{\rm p}$ & $d_{\rm l}$ &  $S$ \\ \hline
PdP   &   PBE  &   NA   & 2.035 &  0.000 &   NA  & 0  \\ \hline
PdP   &   B3LYP  &  (0.000)   & (2.043) &  0.000 &   NA  & 0  \\ 
PdP   &   B3LYP  &  (3.635)   & (2.121) &  0.000 &   NA  & 1  \\ \hline
PdTPP$^a$ &   expt.  &  NA  & 2.009 & 0.000 & NA & 0 \\ 
PdOEP$^b$ &   expt.  &  NA  & 2.018 & 0.000 & NA & 0 \\ \hline
MnP   &   PBE  &  0.000 & 2.003 &  0.000 &   NA  & 3/2 \\
MnP   &   PBE  &  0.495 & 2.051 &  0.000 &   NA  & 5/2 \\ \hline
MnP   &  DFT+U &  0.233 & 2.011 &  0.000 &   NA  & 3/2 \\
MnP   &  DFT+U &  0.000 & 2.090 &  0.000 &   NA  & 5/2 \\ \hline
MnP   &  B3LYP &  0.190 & 2.015  &  0.000 &   NA  & 3/2 \\
MnP   &  B3LYP &  (0.247) & (2.011) &  0.000 &   NA  & 3/2 \\
MnP   &  B3LYP &  0.000  & 2.090 &  0.000 &   NA  & 5/2 \\ 
MnP   &  B3LYP &  (0.000) & (2.089) &  0.000 &   NA  & 5/2 \\ \hline
MnTPP$^c$&  {\rm expt}  &   NA   & 2.085 &  0.000 &   NA  & 5/2 \\ \hline
Mn(H$_2$O)P& PBE &   NA   & 2.013 &  0.024 &  2.39 & 3/2 \\ \hline
Mn(H$_2$O)P& DFT+U &   NA   & 2.108 &  0.129 &  2.34 & 5/2 \\ \hline
MnClP &   PBE  &  0.492 & 2.014 &  0.255 &  2.17 &  1  \\
MnClP &   PBE  &  0.000 & 2.036 &  0.246 &  2.31 &  2  \\ \hline
MnClP &  DFT+U &  0.545 & 2.040 &  0.278 &  2.28 &  1  \\
MnClP &  DFT+U &  0.000 & 2.045 &  0.267 &  2.31 &  2  \\ \hline
MnClTPP$^d$&  {\rm expt}  &   NA   & 2.001 &  0.156 &   2.296  & 2 \\ \hline
\end{tabular}
\caption[]
{\label{table1} \noindent
Spin states and relative energies of metal porphines,
and some intramolecular distances: metal-nitrogen
($d_{\rm N}$), metal-porphine ring ($d_{\rm p}$),
and metal-ligand ($d_{\rm l}$).  Only the ground
state of Mn(H$_2$O)P is tabulated.  Energies and
distances are in units of eV and \AA, respectively.
B3LYP results with/without parantheses are computed using
different basis sets (see text).
\newline
Experimental results in the solid phase: \protect \\
$^a$tetraphenylporphinatopalladium
(II) (PdTPP);\cite{pd_expt1} \protect\\
$^b$octaethylporphinatopalladium (II)
(PdEOP);\cite{pd_expt2} \protect\\
$^c$tetraphenylporphinatomanganese (II)
(MnTPP);\cite{mn_expt1} \protect\\
$^d$tetraphenylporphinatomanganese (III) \protect \\
chloride (MnCTPP).\cite{mn_expt3}
}
\end{table}

Table~\ref{table1} shows that B3LYP predicts a high spin
($S=5/2$) ground state for MnP, more stable by 0.21 (0.25)~eV over the
intermediate spin-state ($S=3/2$).  The values without/with parentheses
are computed using the all electron (TZV/6-311G**//6-311+G/6-31G*)
basis set, or lacv3p** (Mn) and 6-311G** (other atoms), respectively.
They show that neither the high spin-intermediate spin energy splitting, nor
the molecular geometry, is strongly affected by the basis set used.
(We have listed the average of the two Mn-N distances
in the case of quartet MnP, which has a D$_{\rm 2h}$ symmetry.)
The stability of the high spin state is in agreement with experiments on
substituted manganese porphyrins.\cite{mn_expt}  
PBE, on the other hand, incorrectly
predicts that $S=3/2$ is more stable than $S=5/2$ by $\sim 0.50$~eV.
When we set $U=4.2$~eV, $J=1.0$~eV
within the DFT+U approach, the $S=5/2$ spin polarized
state becomes more stable than the $S=3/2$ state by 0.23~eV.
This is sufficiently close to the B3LYP results for our
purpose.\cite{note2,kozlowski}
Since DFT+U predicts that all Mn(II)P and Mn(III)P species examined
in this work are high spin, these ground state predictions
are not affected by spin contamination.\cite{spin_contam}
We emphasize that we use B3LYP as a standard to parameterize our
DFT+U work because it yields the correct high-spin ground
state for MnP;\cite{mn_expt1} as discussed in the introduction, B3LYP
is not universally successful for all transition metal compounds.

With these parameters, the ground state Mn-N distance is predicted to be
$d_{\rm N}=2.090$~\AA, very similar to the B3LYP prediction, and
within 0.005~\AA\, of the experimental value of 2.085~\AA, measured
for the corresponding tetraphenylporphyrin derivative.\cite{mn_expt1}
On the other hand, even when the total spin is constrained to $S=5/2$,
PBE predicts a Mn-N distance of 2.051~\AA, or 0.034~\AA\, smaller
than the experimental value.  Thus, even with a high-spin constraint,
PBE underestimates Mn-N distances in Mn(II)P,
while the DFT+U approach can be parameterized to yield both the correct
spin state and a reasonable metal-porphine geometry.

Attaching a H$_2$O ligand to Mn does not qualitatively
change the above analysis, with PBE still predicting an incorrect
$S=3/2$ spin state and a Mn-N distance substantially smaller than
that predicted using the DFT+U method.

The energy splitting between the high and intermediate spin states is
sensitive to $U$, while the Mn-N distance is less sensitive.  For example,
setting $U=3.8$~eV instead of 4.2~eV yields a 0.12~eV splitting, but
the Mn-N bond length remains essentially unchanged at 2.088~\AA.

In MnPCl, spectroscopic measurements have confirmed that
Mn is in the high spin Mn(III) oxidation state,\cite{mn_expt2} 
in agreement with DFT+U and PBE predictions.  These methods
also predict similar Mn-N bond lengths that overestimate the experimental
value in substituted porphyrins by up to 0.04~\AA.\cite{mn_expt2,mn_expt3}
This discrepancy may be related to the $\sim 0.1$~\AA\,
difference in the predicted and measured out-of-plane Mn displacement.
In the gas phase, the predicted change in Mn-N distance between the
Mn(III) and Mn(II) oxidation states is thus underestimated.  Despite this,
as will be seen, DFT+U predicts a significant increase in this bond length
when Mn(III)P adsorbed on Au(111) is switched to a Mn(II)-like oxidation
state with an electric field.

PBE predicts that PdP is low spin, $S=0$, in agreement with
experiments.\cite{pd_expt3,pd_expt4}  So does DFT+U (not shown).
In general, the $4d$ electrons in second row transition metal ions 
are less localized than $3d$ electrons in first row transition metal
ions, and non-hybrid GGA methods such as PBE appear more reliable
in predicting the ordering of their spin states than is the case
with 3d electron systems.\cite{second_row}

\subsection{Normal-Coordinate Structural Decomposition}

The conformations of the porphine macrocycles in the predicted
structures are further examined using Normal-Coordinate Structural
Decomposition (NSD).\cite{ncd}  NSD has emerged as a useful method
for analyzing deformations of tetrapyrrole
macrocycles in heme proteins and in synthetic proteins,\cite{ncd2} and
such analyses are pertinent to the potential use of transition
metal porphyrins adsorbed on metal electrodes as conformational
switches.\cite{shel1,shel2}
Complete NSD analyses of selected structures from Table~I are given
in Tables S4-S13 of the Supporting Information.  The PBE predicted
structure for PdP and DFT+U predicted ones for MnP and MnClP show
negligible amounts (less than 0.03\AA) of nonplanar deformation, consistent
with the crystallographic data for other transition metal porphines
(see Table S13).  For comparison, highly substituted and very nonplanar
porphyrin macrocycles
exhibit up to 4~\AA\, deformations in the soft ruffling (B$_{\rm 1u}$)
or saddling (B$_{\rm 2u}$) modes.\cite{ncd3}

NSD also provides details of the in-plane deformations present
in the isolated porphine macrocycles.\cite{ncd}   In-plane deformations
are not usually analyzed in any detail because they are naturally intertwined
with out-of-plane deformations in substituted porphyrins,\cite{ncd3}
making it difficult to isolate the former.  However, in the case of
the porphines where the macrocycles are all nominally planar, it is
possible to see the in-plane deformations resulting from the changes
in the size of the metal complexed to porphine.  For example, in
the structures of PdP (PBE), MnClP (DFT+U), and MnP (DFT+U), the
metal-nitrogen distances are 2.035, 2.045, and 2.090~\AA, respectively.
(The X-ray structure of nickel (II) porphine exhibits an even smaller
metal-nitrogen distance of 1.951~\AA\, because of the small size of the
Ni(II) ion; see Table~S13.) The NSD studies
reveal a corresponding increase in the first-order A$_{\rm 1g}$ (breathing)
deformation as the porphine ring expands to accommodate the larger metal ion.
The first order A$_{\rm 1g}$ deformations are +0.15, +0.11, and +0.37~\AA\,
(and -0.16~\AA\, for NiP).  Finally, we note that the
discrepancy between the metal-nitrogen distances predicted for PdP and
measured in PdTPP and PdOEP may be related to the in-plane deformations
that accompany the out-of-plane deformations in the latter, as the
crystal structures of these Pd complexes adopt nonplanar conformations of the
type that are known to shorten the metal-nitrogen bond\cite{shel_para}
(e.g., the large ruffling (B$_{\rm 1u}$) out-of-plane deformation in
PdTPP (see Table~S5)).
 
\subsection{Palladium (II) porphine on Au(111)}

We consider two adsorption geometries: 
a flat adsorption geometry suggested by experiments on substituted
porphyrins\cite{mol_elec1,mol_elec2,mol_elec3,mol_elec4}
(Fig~\ref{fig2}a); and the edgewise
porphyrin stack configuration examined in
Ref.~13,
with successive macrocycles separated by $\sim 5.9$~\AA\, (Fig~\ref{fig2}b).

\subsubsection{Edgewise adsorption}
 
\begin{figure}
\centerline{ \hbox{\epsfxsize=2.90in \epsfbox{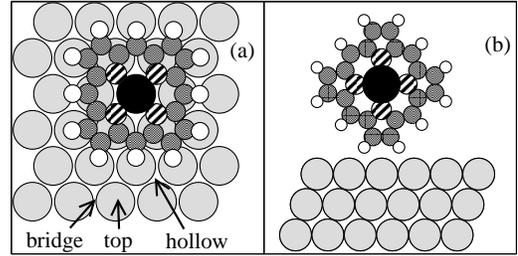}}}
\caption[]
{\label{fig2} \noindent
Palladium porphine on Au(111).  (a) Flat adsorption geometry, with
Pd on the top site.  The top, hollow, and bridge sites are
illustrated.  (b) Edgewise adsorption.  
Striped circle: N; black: Pd; dark grey: C; white: H; light grey: Au.
Periodic boundary conditions are used in all three directions (a vacuum
layer separates the gold surfaces); only the gold atoms in the primitive
simulation cell are shown.  For the geometry in (b), the porphine molecules
are in fact self-assembled in a tube-like stack along the gold surface.
}
\end{figure}

When we attempt to adsorb PdP on Au(111) in the edgewise geometry depicted
in Fig.~\ref{fig2}b using the PBE exchange correlation functional,
we find little or no attraction between the PdP tube and the gold substrate.
In fact, PBE even predicts that the PdP molecules repel each other when
they reside 5.91~\AA\, apart in the self-assembled tube-like geometry
in the {\it absence} of gold.  The formation energy of
this PdP tube is +0.06~eV when using 4$\times$2$\times$1
Brillouin zone sampling.  The small repulsion is partly due
to the fact that PBE underestimates van der Waals attractions.
Surprisingly, we find that the $\Gamma$-point Brillouin zone sampling
used in
Refs.~13 and~14
yields an (unconverged) 0.38~eV repulsion.  This is despite the fact that
there is at most a 0.04~eV dispersion in the PdP valence electron states
near the Fermi level.  Regardless of which PdP reference energy is
used --- the isolated porphine or porphine tube --- we find
no attractive interaction between PdP and Au(111), in contrast
to the $\sim 10$~eV binding energy as previously
reported by Lamoen {\it et al}.\cite{parrin_sur}.

Lamoen {\it et al}.~applied LDA, not the PBE exchange correlation
functional.  They reported that the closest distance between PdP edge
protons and Au atom is 1.78~\AA, suggesting a strong covalent bond between
H and Au.  The large adsorption energy was rationalized by comparing this
system with the interaction of hydrogen molecules on transition metal
surfaces.\cite{parrin_sur}  While LDA is known to predict over-binding
of molecules on metal surfaces, the discrepancy between our predictions
and
Ref.~13
is too large to be explained by the
different treatment of exchange/correlation.  To resolve
this discrepancy, we revisit the edgewise adsorption binding
geometry using LDA.  We apply both the VASP code,\cite{vasp} with 
ultrasoft pseudopotentials and a plane wave basis, and the SeqQuest
code,\cite{seq} which applies norm-conserving pseudopotentials and localized
Gaussian basis sets.  
We use two different DFT packages
to ensure that our predictions are not artifacts of the pseudopotentials
employed.  With both of these DFT codes, our predicted equilibrium
Au lattice constant of 4.078~\AA\, agrees with
Ref.~13.
However, starting with a geometry where two edge PdP protons are 1.8~\AA\,
away from surface Au atoms, PdP experiences large forces and relaxes
away from the substrate.  The optimized distance of closest approach
between PdP protons and Au atoms is 2.36~\AA, and is associated with an
adsorption energy of only 0.25~eV per PdP molecule.  (The relaxed geometries
are listed in Supporting Information section.)
Sampling different adsorption sites by moving the PdP center of
mass in the $x$-$y$ plane results in only small
($\sim 0.03$~eV) variations in $E_{\rm bind}$.

We conclude, based on the results of two very different DFT codes, that
the attraction between Au(111) and edgewise adsorbed PdP is weak, of the
order of 0.25~eV within the LDA approximation, not the $\sim 10$~eV reported
previously.\cite{parrin_sur}  
Note that a DFT work which adopts the edgewise adsorption geometry of
Ref.~13
to study PdP on Al(111) also reported a LDA binding
energy of a small fraction of an~eV.\cite{par_clone}

We emphasize that this 0.25~eV attractive interaction is
computed using LDA.  In the remainder of this work, we will use PBE
or DFT+U based on PBE.  The PBE exchange correlation functional predicts
a negligible binding energy between an edgewise adsorbed PdP tube on Au(111).

\subsubsection{Flat adsorption geometry}

Next, we examine a flat adsorption geometry.  We place the center
of mass of the PdP molecule, namely the Pd atom, atop the
hollow, bridge, and top sites of the Au(111) surface,
and carry out geometry optimization.  $\Gamma$-point sampling and
the 3-layer, 90 atom Au(111) supercell described previously
are used.  Due to the weak interaction between PdP and Au(111),
geometric relaxation along the $z$-direction is slow.  Nevertheless, 
when PdP previously optimized in the gas phase
is initially placed 3.17~\AA\, from the top layer of Au(111) atoms,
the porphine ring eventually relaxes to
a distance $\sim 3.5$~\AA\, above the gold surface.  Here, the
$z$ coordinate of the top layer of the gold substrate is averaged
over all atoms in that layer, and the porphine ring $z$-position
is averaged over all porphine atoms other than Pd.

The binding energies for all three sites are similar, and are
between 0.255~eV and 0.270~eV (Table~\ref{table2}).  They are
consistent with a weak, non site-specific van der Waals interaction
between PdP and Au(111).  There are minimal changes in the
PdP geometry compared to the gas phase system, with the Pd-N distance
expanding by less than 0.01~\AA, and the Pd(II) ion displacing
towards the gold substrate by less than 0.1~\AA.
PBE underestimates van der Waals forces
between PdP and Au.  For a qualitative comparison, our molecular
force field calculation\cite{shel_para} yields a 0.54~eV binding
energy for bare porphine (H$_2$P) molecules adsorbed in a flat
geometry on Au(111).
Increasing the number of gold layers from 3 to 4 increases the PBE
$E_{\rm bind}$ by only 0.005~eV, showing that the calculation
is adequately converged with respect to the system size employed.

\begin{table}\centering
\begin{tabular}{ c c c c c c } \hline
site & $E_{\rm bind}$ & $d_{\rm N}$ & $d_{\rm p}$ & $d_{\rm Au}$ &
$d_{\rm p/Au}$ \\ \hline
top    &  -0.270  & 2.040 & -0.079 & 3.40 & 3.48 \\
bridge &  -0.267  & 2.039 & -0.079 & 3.40 & 3.48 \\
hollow &  -0.255  & 2.039 & -0.064 & 3.39 & 3.45 \\ \hline
\end{tabular}
\caption[]
{\label{table2} \noindent
Flat adsorption geometry PdP binding energies on various Au(111) surface
sites and some intramolecular distances:
metal-nitrogen ($d_{\rm N}$), metal-porphine ring ($d_{\rm p}$),
metal-gold surface ($d_{\rm Au}$), and porphine-gold surface
($d_{\rm p/Au}$).
Calculations are performed using 3-layers of Au atoms and $\Gamma$-point
sampling.  Energies and distances are in units of eV and \AA, respectively.
}
\end{table}

\begin{figure}
\centerline{ \hbox{\epsfxsize=3.00in \epsfbox{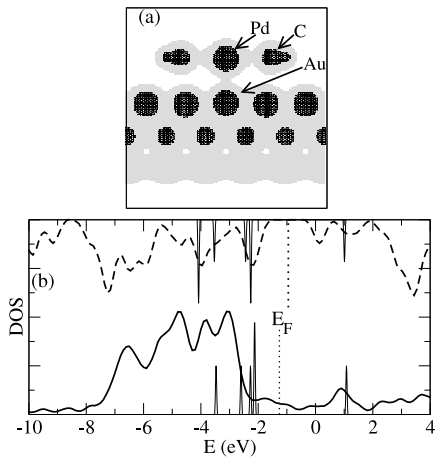}}}
\caption[]
{\label{fig3} \noindent
(a) Distribution of electron density in a $y$-cross sectional plane through
the Pd atom.  Light (dark) shaded regions denote densities of at least 0.1
(1.0) electron/\AA$^3$.
(b) Electronic density of states for palladium porphine (PdP) adsorbed
at the Au(111) top site.   The solid and (inverted) dashed curves
depict the DOS of the PdP-Au complex and of an isolated PdP
molecule, respectively, broadened by gaussians of width 0.1~eV.
The sharp lines depict contributions from orbitals which have
substantial Pd $4d$ character.  The Fermi levels
($E_{\rm F}$) are shown as dotted lines.
}
\end{figure}

The absence of PdP-Au(111) covalent bonding is confirmed by examining
the electron density distributions and the
electronic density of states (DOS).  
The electron densities associated with PdP and the Au surface
presented in Fig.~\ref{fig3}a show minimal overlap.
Figure~\ref{fig3}b superimposes the DOS 
of an isolated PdP (inverted curve in the upper half of the figure)
with that of the adsorbed PdP-Au(111) complex (lower curve).
We align the two sets of DOS by performing
calculations where PdP is moved far from
Au(111) surface, and line up the lowest occupied orbitals,
namely the Pd $4p$ states.
We find that a 6~or~8~\AA\, separation is sufficient to
converge the DOS alignment to the infinite separation limit.
The DOS's show that the Fermi level of the PdP-Au complex
resides within the highest/lowest occupied molecular orbital
gap of PdP, far from PdP $d$-orbital levels.  Thus, we do not expect any
charge transfer between PdP and Au.  Figure~\ref{fig3}(b) also depicts
the orbitals in isolated PdP and the PdP-Au complex which
exhibit substantial Pd $4d$ character.  There are more than 5 such
states because of hybridization between the Pd and nitrogen orbitals.
The highest $4d$-dominated state is unoccupied,
consistent with the $4d^8$ Pd(II) electronic configuration.
Adsorbing PdP on to Au(111) perturbs these Pd $4d$-like
orbitals by only a small fraction of an eV,
and their occupancies do not change.  

To summarize, using the PBE exchange correlation functional,
we find that PdP lies flat on Au(111) surface,
with a binding energy of 0.27~eV.  It does not adsorb
in an edgewise geometry.  
This conclusion is in qualitative
agreement with experimental observation of substituted porphyrins
lying flat on gold surfaces.\cite{mol_elec1,mol_elec2,mol_elec3,mol_elec4}
The interaction is dominated by van der Waals forces
and is not surface site specific.

\subsection{Mn(II) porphine on Au(111) --- PBE predictions}

The adsorption of MnP on Au(111) is qualitatively different from that
of PdP.  First, we consider PBE predictions using 4 Au layers and
$\Gamma$-point sampling.  While PBE does not predict the correct
MnP high spin ground state, we will show that
it yields qualitatively correct trends for MnP adsorbed on
Au(111) in the absence of an electric field.  Since DFT+U
calculations are more expensive than PBE, it is
also more convenient to use the latter for convergence tests
with respect to system size.

Table~\ref{table3} shows that the PBE binding energies for
MnP on Au(111) are significantly larger than for PdP, and they are
more site-specific.  The top, bridge, and hollow site
$E_{\rm bind}$ differ by up to 0.12~eV.  The optimized geometries 
indicate that Mn ion is displaced out of the porphine plane towards the
gold surface by more than 0.2~\AA.
On the other hand, the porphine ring remains $\sim 3.5$~\AA\, from the
gold surface, demonstrating that the stronger attraction arises
from the Mn ion interacting with the gold substrate.  
The magnitude of the out-of-plane Mn displacement towards Au(111)
correlates with increasing binding energy as the adsorption site varies.
When a H$_2$O ligand is bound to the Mn ion at the top site,
the Mn ion becomes almost coplanar with the porphine ring, and
$E_{\rm bind}$ is slightly reduced, to a value almost identical
to that of the unligated MnP at the hollow site.  In contrast, an
isolated MnPCl is already in the Mn(III) oxidiation state,
and its adsorption energy is drastically reduced.

\begin{table}\centering
\begin{tabular}{ c c c c c c c } \hline
site & $E_{\rm bind}$ & $m$ & $d_{\rm N}$ & $d_{\rm p}$ &
$d_{\rm Au}$ & $d_{\rm p/Au}$ \\ \hline
top         &  -0.800  & 3.77 & 2.010 & -0.224 & 3.27 & 3.49 \\
bridge      &  -0.740  & 3.70 & 2.011 & -0.216 & 3.28 & 3.50 \\ 
hollow      &  -0.680  & 3.68 & 2.014 & -0.205 & 3.29 & 3.50 \\ \hline
top(H$_2$O) &  -0.680  & 3.76 & 2.015 & +0.005 & 3.55 & 3.54 \\ 
top(Cl$^-$) &  -0.179  & 3.74 & 2.028 & +0.242 & 3.74 & 3.50 \\ \hline
\end{tabular}
\caption[]
{\label{table3} \noindent
Flat adsorption geometry binding energies (eV), total spin magnetic moment in
the supercell ($\mu_{\rm B}$), and geometries of MnP, Mn(H$_2$O)P,
and MnClP adsorbed at various Au(111) sites.
Distances are in \AA.  The PBE exchange correlation functional,
4 layers of Au atoms, and $\Gamma$-point Brillouin zone sampling
are applied in these calculations.
The symbols are described in the caption to Table~\ref{table2}.
}
\end{table}

Table~\ref{table4} shows that 4 layers of gold atoms and $\Gamma$-point
sampling converge the binding energy to within $\sim 0.09$~eV.
$E_{\rm bind}$ does not convergence monotonically with system size,
and depends on the degeneracy of the highest unoccupied orbitals of the
Au(111) slab.  Using $\Gamma$-point sampling, the same Fermi level degeneracy
repeats itself whenever 3 layers of gold is added, and, $E_{\rm bind}$
are almost identical for the 3- and 6-layer gold models.
With more $k$-point sampling in the $x$-$y$ directions, better convergence
is obtained.  We will adopt the 4-layer system with $\Gamma$-point sampling
as a compromise between accuracy and computational convenience.
Table~\ref{table4} emphasizes the importance of proper Brillouin zone sampling
when treating adsorption of transition metal embedded molecules on
metal surfaces.  It suggests that using a finite sized, gas phase cluster
geometry to represent the gold substrate, which is always
limited to $\Gamma$-point sampling and results in an
electronic {\it insulator}, may be especially problematic.

The spin magnetic moment ($m$) of the simulation cell is also
converged to $\sim 0.2$~$\mu_{\rm B}$ with 4 layers of Au.
We have adopted the convention, frequently used in the DFT+U
literature,\cite{ldau0,ldau2} that reports the spin magnetic moment $m$ as
the difference in occupation numbers between up-spin and down-spin
orbitals, in units of $\mu_{\rm B}$.
The magnetic moment changes from the PBE prediction of
$m=3$$\mu_{\rm B}$ for isolated MnP to $M = 3.8$$\mu_{\rm B}$ for
the supercell containing the adsorbed molecule.
Decomposition of spin densities on atomic centers show
that the magnetic moment is localized on Mn $3d$ orbitals.
Note that, within the framework of spin-polarized density functional theory,
adding an electron to the periodically replicated gold substrate
contributes zero net spin to the entire system.

\begin{table}\centering
\begin{tabular}{ c c c c } \hline
Au layers & $k$-point & $E_{\rm bind}$ & $m$ \\ \hline
3 & $1\times 1\times 1$ & -0.869  & 3.83 \\
3 & $2\times 2\times 1$ & -0.688  & 3.19 \\
3 & $3\times 3\times 1$ & -0.727  & 3.58 \\
4 & $1\times 1\times 1$ & -0.800  & 3.77 \\
4 & $2\times 2\times 1$ & -0.783  & 3.79 \\
5 & $1\times 1\times 1$ & -0.693  & 3.48 \\
5 & $2\times 2\times 1$ & -0.786  & 3.57 \\
6 & $1\times 1\times 1$ & -0.868  & 3.80 \\ \hline
\end{tabular}
\caption[]
{\label{table4} \noindent
Convergence of MnP binding energy (eV) and spin magnetic moments ($\mu_{\rm B}$)
at the top site of Au(111) as the number of Au layers and Brillouin zone
sampling vary.
}
\end{table}

In this work, we have adopted isolated (``gas phase'') Mn(II)P as
the reference to compare with MnP adsorbed on Au(111).
Although the Mn(II) oxidation state is generally less
stable in porphyrins than Mn(III), properties of stable substituted Mn(II)
porphyrins have been measured,\cite{mn_expt1} and electrochemically induced
transitions between Mn(II) and Mn(III) are readily
realizable.\cite{porph_review,mn_expt2} 
In reality, a gold electrode represents an infinite electron reservoir;
when a non-ligated MnP adsorbs on gold electrode, the final spin state,
charge configuration, and molecular geometry will not be affected by
the initial MnP charge state, although the binding energy measured from
the Mn(III)P reference state would be different.

\subsection{Mn(II) porphine on Au(111) --- DFT+U predictions}

In this section, we apply the DFT+U technique to examine MnP
adsorption at the top site of Au(111).  We use 4 layers of Au atoms with
$\Gamma$-point sampling, previously shown to yield
converged $E_{\rm bind}$ from PBE calculations.  
DFT+U consistently predicts a larger $E_{\rm bind}$ for the top site
than the bridge site, just like the PBE exchange correlation functional.

\begin{figure}
\centerline{\hbox{\epsfxsize=3.00 in \epsfbox{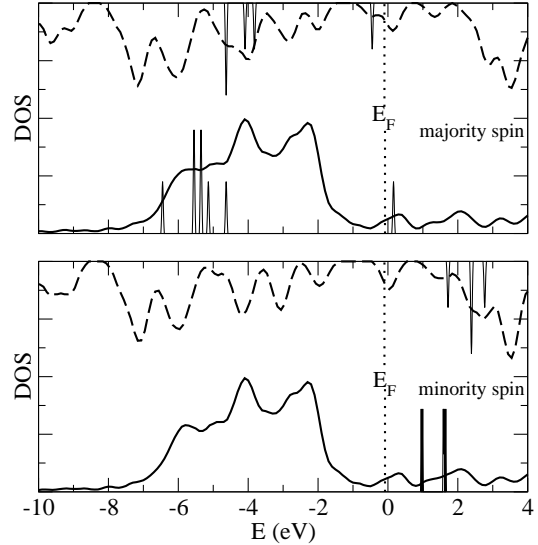}}}
\caption[]
{\label{fig4} \noindent
Same as Fig.~\ref{fig3}(b), but for MnP computed using the DFT+U
exchange correlation functional.  The upper and lower panels
refer to the majority and minority spin channels, respectively.
The sharp lines are states with substantial Mn $3d$ character.
}
\end{figure}

Figure~\ref{fig4} depicts the electronic density of states
in the two spin channels of both isolated MnP and the MnP-Au(111) complex.  
As already shown in Table~\ref{table1}, DFT+U (unlike PBE) yields a
$m=5$$\mu_{\rm B}$ ($S=5/2$) MnP ground state, in agreement with
experiments for substituted MnP.  All majority spin orbitals with
substantial Mn $3d$ character are occupied in isolated MnP, while
no such minority spin orbitals are occupied.

Adsorption of MnP on Au(111) leads to striking changes in the
spectra of orbitals with substantial Mn~3$d$ character,
which are perturbed by up to 2~eV.   The highest
occupied majority spin Mn-3$d$ orbital
in isolated MnP is pushed into the conduction band.   Overall,
the MnP-Au(111) complex exhibits a magnetic moment of $m=3.9$$\mu_{\rm B}$,
similar to PBE results.  While both DFT+U and PBE predict
similar adsorption geometries and binding energies (see
Tables~\ref{table3} and~\ref{table5}), 
their respective Mn $3d$-like orbitals (not shown for PBE)
differ in energies by up to 2~eV.

\begin{table}\centering
\begin{tabular}{ c c c c c c c c } \hline
species & field & $E_{\rm bind}$ & $m$ & $d_{\rm N}$ &
$d_{\rm p}$ & $d_{\rm Au}$ & $d_{\rm p/Au}$ \\ \hline
MnP       & 0.000 &  -0.686  & 3.87 & 2.022 & -0.247 & 3.25 & 3.50 \\
MnP       & 0.700 &  -0.145  & 4.88 & 2.156 & -0.540 & 2.94 & 3.48 \\ 
MnP$^*$   & 0.700 &  -0.152  & 3.22 & 2.011 & -0.125 & 3.39 & 3.52 \\ \hline
Mn(H$_2$O)P & 0.000 &  -0.737  & 3.91 & 2.020 & +0.005 & 3.55 & 3.55 \\ 
Mn(Cl)P   & 0.000 &  -0.106  & 3.94 & 2.040 & +0.187 & 3.68 & 3.49 \\  \hline
\end{tabular}
\caption[]
{\label{table5} \noindent
The binding energies (eV), the total spin magnetic moment in the supercell
($\mu_{\rm B}$), and molecular geometries of MnP and Mn(H$_2$O)P in
zero and 0.7 V/\AA\, applied field.  Distances are in \AA.
The DFT+U method is applied except for the case marked with the asterisk,
which uses the PBE functional.  The symbols and system size are
described in the caption to Table~\ref{table3}.
}
\end{table}

\begin{figure}
\centerline{\hbox{\epsfxsize=3.00 in \epsfbox{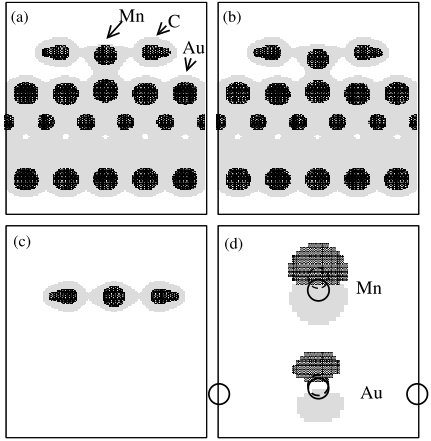}}}
\caption[]
{\label{fig5} \noindent
Distributions of electron density in a $y$-cross sectional plane through
the Mn atom, computed using the DFT+U method.  (a) No applied field;
(b) 0.7~eV applied field; (c) isolated MnP, zero field; (d)
electron density difference between (a) and (b).
In panels (a), (b), and (c), light (dark) shaded regions denote densities
of at least 0.1 (1.0) electron/\AA$^3$.
In (d), the lighter (darker) shades depict regions
of electron loss (gain) by at least 0.2 electron/\AA$^3$ upon applying
the field.  The circles drawn with solid (dashed) lines donote the
Mn and Au atom positions with (without) applied field.  Note that
these atoms and their electron clouds move in opposite directions.
}
\end{figure}

Figure~\ref{fig5}a illustrates the charge densities in a cross
sectional $y$ plane through the Mn ion and the first layer Au atoms.
In contrast to PdP (Fig.~\ref{fig3}), there is significant
overlap of charge densities between Mn and the top site
Au atom closest to it.  On the other hand,
the $\pi$-electrons on carbon atoms of MnP clearly do
not interact with Au(111).  Comparing the
charge density of isolated MnP (Fig.~\ref{fig5}c) with that
of adsorbed MnP, the top side, out-of-porphine-plane electron density in
the isolated MnP is diverted to the substrate side, where
it overlaps with gold orbitals.

To determine the extent of electron transfer between MnP and Au(111),
we make a cut through the narrowest region of the funnel-shaped electron
density distribution, and assign charges to the two species accordingly.
We find there is at most a transfer of 0.2~electron from MnP to the
substrate, despite the drastic changes in the DOS and the spin multiplicity
from their gas phase values.  The binding between MnP and Au(111)
is thus best described as covalent or metallic.

\subsection{Applying an electric potential}

Given that the MnP-Au(111) interaction is associated with
partial electron transfer, we expect that
changing the electric potential on the gold substrate
will strongly influence the adsorption behavior.  An electric
field which favors transfer of electrons
from MnP to Au can strengthen the MnP-Au interaction.  An electric
field which favors electron transfer back into
MnP $3d$ orbitals should cause the Mn(III)-like behavior to revert
back to Mn(II).

Applying electric fields within plane wave density functional
theory calculations has been formulated in the literature
and implemented into the VASP code.\cite{feib}
We apply a field of 0.7~V/\AA\, to isolated MnP, a 4-layer
gold slab, and the adsorbed MnP on 4-layers of Au atoms.
$E_{\rm bind}$ is obtained as the energy difference between the adsorbed
and isolated systems.  $\Gamma$-point sampling is used throughout.

Table~\ref{table5} shows that a 0.7~V/\AA\, electric field
strongly reduces the binding energy and changes the
magnetic moment to almost its isolated MnP value: $m=4.89$$\mu_{\rm B}$
according to DFT+U predictions, compared with $m=5$$\mu_{\rm B}$
in isolated MnP.
The Mn-N distance, $d_{\rm N}$, also increases from the zero field
value of 2.02~\AA\, to 2.15\AA, which is even larger than the value of
$d_{\rm N}=2.090$\AA\, predicted in isolated MnP.  
The PBE predictions for molecular geometry
and spin states are substantially
different; this reflects the inability of PBE to predict
the correct spin state and molecular geometry for isolated MnP.

The larger Mn-N distance in adsorbed MnP in an electric field compared
to isolated MnP is due to the significant Mn ion displacement out of the
porphine plane towards the gold substrate, and not to a change
in the porphine conformation, which remains essentially planar.
This is demonstrated by the similar core sizes for isolated MnP (2.090~\AA)
and for MnP adsorbed on Au(111) with an electric field (2.087~\AA) (the
core size is defined as the radius of a cylinder which can fit through
the porphine hole).  NSD analysis (supporting information) also
shows that the A$_{\rm 1g}$ (in-plane) deformation of MnP on Au(111)
in the absence of an applied field is consistent with a
Mn(III)-like species, with a A$_{\rm 1g}$ deformation
of +0.08~\AA\, compared to +0.11~\AA\, predicted for isolated MnClP.
As expected, with an applied electric field of 0.7~V/\AA, the first order
A$_{\rm 1g}$ deformation of +0.31~\AA\, is similar to that predicted for 
isolated MnP (+0.37~\AA).

Figure~\ref{fig5} illustrates the effect of the applied
field on the charge density.  The field evidently
repopulates the Mn $3d_{x^2-y^2}$-like orbital and restores
electron density to the top side of the MnP molecule.
There still appears substantial electron cloud overlap
between MnP and the gold surface despite the small $E_{\rm bind}$.
Figure~\ref{fig5}d depicts the induced charge differential.
Note that the field causes both the surface top site Au atom
and the Mn ion to move towards the gold substrate, whereas the
electron densities on both atoms are displaced in the opposite direction.

\begin{figure}
\centerline{\hbox{\epsfxsize=2.90 in \epsfbox{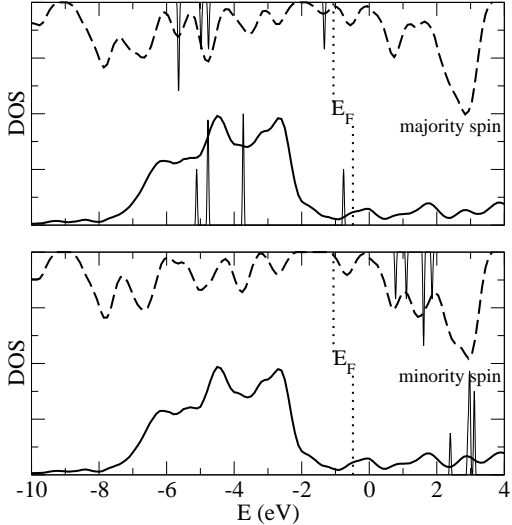}}}
\caption[]
{\label{fig6} \noindent
Same as Fig.~\ref{fig4}, for MnP computed using the DFT+U method with
an applied electric field of 0.7~eV/\AA.
}
\end{figure}

Figure~\ref{fig6} plots the electronic DOS predicted by DFT+U
in the applied field.  Compared with the zero field case (Fig.~\ref{fig4}),
the occupied states are shifted to lower energies relative
to the Fermi level.  In particular, the highest lying orbital
with substantial Mn $3d$ character is brought below the Fermi
level in the MnP-Au(111) complex, and is occupied by electrons.  This
is consistent with the field induced change in the total magnetic
moment.  Nevertheless, if we demarcate the charge density at the
narrowest region of the electron cloud between Au and Mn, once again
we find a small, $\sim 0.1$ electron difference between the total charge
on MnP and Au with or without the electric field.  Thus,
while the charge and spin state of the Mn ion differ, the locus
of the total electron density is not strongly affected.

The 4-layer bare gold slab is stable in the 0.7~V/\AA\, field, and
the field-induced change in its energy is proportional to the square of
the field magnitude.  We find that smaller, 0.525~V/\AA\, and 0.35~V\AA\,
electric fields lead to similar behavior described for MnP adsorbed on
gold in a 0.7V/A field, although the Mn-N bonds becomes shorter (2.109
and 2.102~\AA, respectively).
Due to the technical difficulty of converging a DFT+U calculation in a 
system with spin polarization, a zero band gap, and an electric field,
we have not located the minimal field needed to trigger this switching.
This will be investigated in a future work.  These calculations
are performed in vacuum, and we have not attempted to relate
the field strength to an electrostatic potential.  In electrochemical cells,
where the molecules and electrodes are immersed in electrolytes, the
voltage difference required to reduce Mn(III) porphyrins to Mn(II) 
is much smaller: about 0.4~Volt in water,\cite{mn_expt2} and on the
order of 0.2-0.3~Volt in organic solvents.\cite{porph_review}

\section{Conclusions}
 
In this work, we use a combination of density functional theory (DFT)
in the generalized gradient approximation (GGA) and the DFT+U technique
to study the adsorption of transition metal porphine molecules
on atomistically flat Au(111) substrates.  The PBE exchange correlation
functional is adequate for treating palladium porphine (PdP).  We
find that PdP preferentially adsorbs flat on Au(111) surfaces, in
agreement with experimental observations of substituted porphyrins
lying flat on gold surfaces.\cite{mol_elec1,mol_elec2,mol_elec3,mol_elec4}
The binding energy is 0.27~eV, irrespective
of the adsorption site.  There is no charge transfer or covalent
bonding between PdP and Au, and the interaction is
predominantly dispersive.  In contrast to a previous DFT
calculation,\cite{parrin_sur} we find that a self-assembled tube-like
geometry with the PdP adsorbing edgewise on Au(111) is not
energetically favorable.  Using multiple DFT codes and types of
pseudopotentials, we find that the LDA binding energy for
this edgewise adsorption geometry is 0.25~eV, not $\sim 10$~eV
as previously predicted.\cite{parrin_sur}   The PBE exchange correlation
functional predicts little or no binding.  Furthermore, an accurate
binding energy for this geometry requires more extensive
Brillouin zone sampling than has been attempted in the literature.

The behavior of manganese porphine (MnP) is qualitatively different.  The PBE
exchange correlation functional does not reproduce the correct
ground state (high spin $S=5/2$ state) or the experimental
Mn-nitrogen distance.  However, PBE can be augmented with the
DFT+U technique, which can be parameterized to yield the correct ordering
of spin multiplicity.  The DFT+U ground state geometry exhibits
a Mn-N distance in good agreement with experiments and B3LYP predictions.

MnP selectively binds to the top site of Au(111).  When we use
Mn(II)P as the reference state, the binding energy of MnP on Au(111)
is significantly larger than for PdP.  The adsorbed MnP exhibits
partial charge transfer from orbitals with substantial Mn $3p$
character to Au(111) states.  The overall magnetic moment
also changes from $m=5$$\mu_{\rm B}$ ($S=5/2$) to $m\approx4$$\mu_{\rm B}$
($S=2$).  In other words, Mn takes on a +3 oxidiation state character.

This MnP-Au(111) interaction can be partially
reversed by applying an electric field with the appropriate polarity.
A sufficiently large field causes the Mn ion to revert back to
a Mn(II)-like spin state, resulting in a significant, up to 0.13~\AA,
increase in the Mn-N distance within the porphine ring.  This suggests
that appropriately substituted Mn porphyrins deposited on gold
electrodes may be useful for demonstrating electric field triggered
conformational changes that are potentially pertinent to harvesting
nanomechanical work and selective ligand binding.\cite{shel1}

In summary, we have successfully applied the DFT+U technique
to transition metal ions in a condensed phase environment with
fully periodic boundary conditions.
With this approach, the energy difference between the spin states
can be parameterized using not just
B3LYP, which gives qualitatively correct spin orderings for many
manganese complexes, but experimental data, quantum chemistry
methods, and quantum Monte Carlo calculations as well.
As such, this versatile technique can potentially be used
for general, accurate, all DFT-based treatment of transition metal ions in
aqueous systems, water-material interfaces, and biological
environments.\cite{klein,water}
 
\section*{Acknowledgement}
 
We thank John Shelnutt, Peter Feibelman, Rick Muller, Richard Martin,
Kevin Zavadil, and Donald Truhlar for useful suggestions.  This work was
supported by the Department of Energy under Contract DE-AC04-94AL85000.
Sandia is a multiprogram laboratory operated by Sandia Corporation, a
Lockheed Martin Company, for the U.S.~Department of Energy.
V.S.B. acknowledges supercomputer time from the National Energy Research
Scientific Computing (NERSC) center and financial support from the National
Science Foundation (NSF) Career Program Award CHE \# 0345984, and the NSF
Nanoscale Exploratory Research (NER) Award ECS \# 0404191. 

\section*{Supporting Information Available:}
A description of the DFT+U implementation,
LDA predicted coordinates of PdP, Au(111) slab, and PdP adsorbed
edgewise on Au(111), as well as Normal-Coordinate Structure Decomposition
(NSD) analyses for some computed structures and crystal
structures, are available free of charge via the
Internet at http://pubs.acs.org.


\begin{references}
 
\bibitem{porph_review}
Kadish,~K.~M.; Smith,~K.~M.; Guilard,~R. {\it The Porphyrin Handbook}
(Academic Press, San Diego, 2000).

\bibitem{sensor} 
Oni,~J.; Diab,~N.; Radtka,~I.; Schuhmann,~W.  {\it Electrochimica Acta}.
{\bf 2003}, {\it 48}, 3349.

\bibitem{memory}
Gryko,~D.~T.; Clausen,~C.; Roth,~K.~M.; Dontha,~N.; Bocian,~D.~F.; Kuhr,~W.~G.;
Lindsey,~J.~S.  {\it J.~Org.~Chem.} {\bf 2002}, {\it 65}, 7345.

\bibitem{light1}
Imahori,~H.; Norieda,~H.; Yamada,~H.; Nishimura,~Y.; Yamazaki,~I.; Sakata,~Y.;
Fukuzumi,~S.  {\it J.~Am.~Chem.~Soc.} {\bf 2001}, {\it 123}, 100;
Nomoto,~A.; Kobuke,~Y.  {\it Chem.~Commun.} {\bf 2002}, 1104.

\bibitem{light2}
Gust,~D.; Moore,~T.~A.; Moore,~A.~L.  {\it Acc.~Chem.~Res.} {\bf 2001},
{\it 34}, 40.

\bibitem{mol_elec1}
He,~Y.; Ye,~T.; Borguet,~E.  {\it J.~Am.~Chem.~Soc.} {\bf 2002}, {\it 124},
11964.

\bibitem{nature}
Yokoyama,~T.; Yokoyama,~S.; Kamikado,~T.; Okuno,~Y.; Mashiko,~S.
{\it Nature} {\bf 2001}, {\it 413}, 619.

\bibitem{sol1}
Vesper,~B.~J.; Salaita,~K.; Zong,~H.; Mirkin,~C.~A.; Barrett,~A.~G.~M.;
Hoffman,~B.~M. {\it J.~Am. Chem.~Soc.} {\bf 2004}, {\it 126}, 16653.

\bibitem{sol2}
Yoshimoto,~S.; Inukai,~J.; Tada,~A.; Abe,~T.; Morimoto,~T.; Osuka,~A.;
Furuta,~H.; Itaya,~K. {\it J.~Phys. Chem.~B} {\bf 2004}, {\it 108}, 1948.

\bibitem{mol_elec2}
Kunitake,~M.; Akiba,~U.; Batina,~N.; Itaya,~K. {\it Langmuir} {\bf 1997},
{\it 13}, 1607; Kunitake,~M.; Batina,~N.; Itaya,~K. {\it Langmuir.}, {\bf 1995},
{\it 11}, 2337.

\bibitem{mol_elec3}
Hipps,~K.~W.; Scudiero,~L.; Barlow,~D.~E.; Cooke,~M.~P. {\it J.~Am.~Chem.~Soc.},
{\bf 2002}, {\it 124}, 2126.

\bibitem{mol_elec4}
Yokoyama,~T.; Yokoyama,~S.; Kamikado,~T.; Mashiko,~S.
{\it J.~Chem.~Phys.} {\bf 2001}, {\it 115}, 3814.
 
\bibitem{parrin_sur}
Lamoen,~D.; Ballone,~P.; Parrinello,~M.  {\it Phys.~Rev.~B} {\bf 1996},
{\it 54}, 5097.
 
\bibitem{par_clone}
Picozzi,~S.; Pecchia,~A.; Cheorghe,~M.; Di Carlo,~A.; Lugli,~P.; Delley,~B.;
Elstner,~M.  {\it Sur.~Sci.} {\bf 2004}, {\it 566}, 628.
 
\bibitem{shaik}
Shaik,~S.; de Visser,~S.~P.; Kumar,~D. {\it J.~Am.~Chem.~Soc.} {\bf 2004},
{\it 126}, 11746.

\bibitem{shel1}
Song,~Y.-J.; Haddad,~R.~E.; Jia,~S.-L.; Hok,~S.; Olmstead,~M.~M.; Nurco,~D.~J.;
Schore,~N.~E.; Zhang,~J.; Ma,~J.-G.; Smith,~K.~M.; Gazeau,~S.; P\'{e}caut,~J.;
Marchon,~J.-C.; Medforth,~C.~J.; Shelnutt,~J.~A.  {\it J.~Am.~Chem.~Soc.}
{\bf 2005}, {\it 127}, 1179.
 
\bibitem{shel2}
Gazeau,~S.; P\'{e}caut,~J.; Marchon,~J.-C.  {\it Chem. Commun.} {\bf 2001},
1644.

\bibitem{shel3}
Mazzanti,~M.; Marchon,~J.-C.; Shang,~M.~Y.; Scheidt,~W.~R.; Jia,~S.-L.;
Shelnutt,~J.~A. {\it J.~Am. Chem.~Soc.} {\bf 1997}, {\it 119}, 12400.

\bibitem{au_expt}
Barth,~J.~V.; Brune,~H.; Ertl,~G.; Behm,~R.~J.  {\it Phys.~Rev.~B} {\bf 1990},
{\it 42}, 9307.

\bibitem{stack_expt}
Burghard,~M.; Fischer,~C.~M.; Schmelzer,~M.; Roth,~S.; Hanack,~M.;
G\"{o}pel,~W.  {\it Chem.~Mater.} {\bf 1995}, {\it 7}, 2104.
 
\bibitem{pbe}
Perdew,~J.~P.; Burke,~K.; Ernzerhof,~M.  {\it Phys.~Rev.~Lett.} {\bf 1996}, 
{\it 77}, 3865.

\bibitem{moroni}
Moroni,~E.~G.; Kresse,~G.; Hafner,~J.; Furthm\"{u}ller,~J.  {\it Phys.~Rev.~B}
{\bf 1997}, {\it 56}, 15629.

\bibitem{selloni}
Di~Felice,~R.; Selloni,~A.  {\it J.~Chem.~Phys.} {\bf 2004}, {\it 120}, 4906.
 
\bibitem{pw91}
Perdew,~J.~P.; Wang,~Y.  {\it Phys.~Rev.~B} {\bf 1992}, {\it 45}, 13244.

\bibitem{second_row}
Here the distinction between 1st and 2nd row transition metal compounds
specifcally refers to the PBE deficiency in treating the coulomb repulsion
between electrons strongly localized at the same site.
For example, in Mott-Hubbard insulators, non-hybrid DFT have been described
as deficient in this respect for $f$-electron compounds, high-$T_c$
cuprates and 3$d$ (but not 4$d$) oxides (see
Ref.~54, pp.~205103-4).

\bibitem{b3lyp}
Becke,~A.~D. {\it J.~Chem.~Phys.} {\bf 1993}, {\it 98}, 1372;
Becke,~A.~D. {\it J.~Chem.~Phys.}, {\bf 1993}, {\it 98}, 5648;
Lee,~C.~T.; Yang,~W.~T.; Parr,~R.~G.  {\it Phys.~Rev.~B} {\bf 1988}, {\it 37},
785.

\bibitem{cost}
See http://cms.mpi.univie.ac.at/vasp/ for preliminary benchmarks obtained
using VASP version~5.

\bibitem{metal}
Hybrid exchange correlation functionals with unscreened, long range exchange
terms may converge slowly with system size for metals, and may be
unsuited for treating metal surfaces. (Martin,~R.~L., private communications).

\bibitem{gga}
See, e.g., Singh,~D.~J.; Pickett,~W.~E.; Krakauer,~H.  {\it Phys.~Rev.~B}
{\bf 1991}, {\it 43}, 11628.

\bibitem{reiher1}
Reiher,~M.; Salomon,~O.; Hess,~B.~A.  {\it Theor.~Chem.~Acc.} {\bf 2001},
{\it 107}, 48.
 
\bibitem{reiher2}
Reiher,~M.  {\it Inorg.~Chem.} {\bf 2002}, {\it 41}, 6928.
 
\bibitem{leung01}
Leung,~K.  {\it Phys.~Rev.~B} {\bf 2002}, {\it 65}, 012102.
 
\bibitem{klein}
Ivanov,~I.; Klein,~M.~L.  {\it J.~Am.~Chem.~Soc.} {\bf 2005}, {\it 127}, 4010.

\bibitem{ed1}
Lundberg,~M.; Siegbahn,~P.~E.~M. {\it Phys. Chem. Chem. Phys.}
{\bf 2004}, {\it 6}, 4772.

\bibitem{ed2}
Lundberg,~M.; Siegbahn,~P.~E.~M.  {\it J.~Comp. Chem.} {\bf 2005}, {\it 26},
661.

\bibitem{ed3}
Koizumi,~K.; Shoji,~M.; Nishiyama,~Y.~M.; Maruno,~Y.;
Kitagawa,~Y.; Soda,~K.; Yamanaka,~S.; Okumura,~M.; Yamaguchi,~K.
{\it Int.~J. Quan. Chem.} {\bf 2004}, {\it 100}, 943.

\bibitem{isobe}
Isobe,~H.; Soda,~T.; Kitagawa,~Y.; Takano,~Y.; Kawakami,~T.; Shyoshioka,~Y.;
Yamaguchi,~K. {\it Int. J. Quantum Chem.}, {\bf 2001}, {\it 85}, 34.

\bibitem{devisser}
de Visser,~S.~P.; Ogliaro,~F.; Gross,~Z., Shaik,~S. {\it Chem.~Eur.~J.}
{\bf 2001}, {\it 7}, 4954.

\bibitem{khav}
Khavrutskii,~I.V.; Musaev,~D.~G.; Morokuma,~K. {\it Inorg. Chem.}, 
{\bf 2003}, {\it 42}, 2606.

\bibitem{prelim}
Prodan,~I.~D.; Scuseria,~G.~E.; Sordo,~J.~A.; Kudin,~K.~N.; Martin,~R.~L.
{\it J.~Chem.~Phys.} {\bf 2005}, {\it 123}, 014703, and references therein.

\bibitem{cpl}
Ricca,~A.; Bauschlicher,~C.~W.  {\it Chem.~Phys.~Lett.}  {\bf 1995},
{\it 245}, 150.

\bibitem{truhlar}
Schultz,~N.~E.; Zhao,~Y.; Truhlar,~D.~G.  {\it J.~Phys.~Chem.~A}.  {\bf 2005},
{\it 109}, 4388, and references therein.

\bibitem{ed4}
Chai,~J.-F.; Zhu,~H.-P.; Stuckl,~C.~A.; Roesky,~H.~W.; Magull,~J.; Bencini,~A.;
Caneschi,~A.; Gatteschi,~D. {\it J.~Am. Chem. Soc.} {\bf  2005}, {\it 127},
9201.

\bibitem{spro}
Sproviero,~E.~M.; Gascon,~J.~A.; McEvoy,~J.~P.; Brudvig,~G.~W.; Batista,~V.~S.
{\it J. Biol. Inorg. Chem.}, {\bf 2005} (in press).

\bibitem{ghosh_rev}  
Ghosh,~A.; Taylor,~P.~R. {\it Curr.~Opin.~Chem.~Biol.} {\bf 2003}, {\it 7}, 113;
Ghosh,~A.; Persson,~B.~J.; Taylor,~P.~R.  {\it J. Biol. Inorg. Chem.}
{\bf 2003}, {\it 8}, 507.

\bibitem{ghosh4}
Ghosh, A.; Taylor, P.~R.  {\it J. Chem. Theory. Comput.} {\bf 2005}, {\it 1},
597.

\bibitem{qmc}
Aspuru-Guzik,~A.; el Akramine,~O.; Grossman,~J.~C.; Lester,~W.~A.
{\it J.~Chem.~Phys.} {\bf 2004}, {\it 120}, 3049.

\bibitem{ldau0}
Anisimov,~V.~I.; Zaanen,~J.; Andersen,~O.~K.   {\it Phys.~Rev.~B}, {\bf 1991},
{\it 44}, 943.

\bibitem{ldau1}
Liechtenstein,~A.~I.; Anisimov,~A.~I.; Zaanen,~J. {\it Phys.~Rev.~B}
{\bf 1995}, {\it 52}, 5467.
 
\bibitem{nio}
Anisimov,~V.~I.; Kuiper,~P.; Nordgren,~J.  {\it Phys.~Rev.~B} {\bf 1994},
{\it 50}, 8257.

\bibitem{lacoo3}
Nekrasov,~I.~A.; Streltsov,~S.~V.; Korotin,~M.~A.; Anisimov,~V.~I.
{\it Phys.~Rev.~B} {\bf 2003}, {\it 68}, 235113.

\bibitem{pickett}
Pickett,~W.~E.; Erwin,~S.~C.; Ethridge,~E.~C. {\it Phys.~Rev.~B.} {\bf 1998},
{\it 58}, 1201.

\bibitem{rohrbach}
Rohrbach,~A.; Hafner,~J.; Kresse,~G. {\it Phys.~Rev.~B} {\bf 2004}, {\it 69},
075413.

\bibitem{singh}
Johannes,~M.~D.; Mazin,~I.~I.; Singh,~D.~J. {\it Phys.~Rev.~B}, {\bf 2005},
{\it 71}, 205103, and references therein.

\bibitem{water}
Asthagiri,~D.; Pratt,~L.~R.; Paulaitis,~M.~E.; Rempe,~S.~B.
{\it J.~Am.~Chem.~Soc.} {\bf 2004}, {\it 126}, 1285.

\bibitem{mn_expt}
Miller,~J.~C.; Sharp,~R.~R. {\it J.~Phys.~Chem.~A} {\bf 2000}, {\it 104}, 4889,
and references therein.

\bibitem{mn_expt1}
Kirner,~J.~F.; Reed,~C.~A.; Scheidt,~W.~R.  {\it J.~Am.~Chem.~Soc.} {\bf 1977},
{\it 99}, 1093.

\bibitem{siegbahn}
Siegbahn,~P.~E.~M.  {\it Current Opin. Chem. Biol.} {\bf 2002}, {\it 6}, 227.
 
\bibitem{spectra}
We have not conducted detailed studies of the MnP optical spectra or excited
states predicted by the DFT+U technique.  This will be the subject of future
work.

\bibitem{vasp}
Kresse,~G.; Furthm\"{u}ller,~J.  {\it Phys.~Rev.~B} {\bf 1996}, {\it 54}, 11169;
{\it Comput.~Mater.~Sci.} {\bf 1996}, {\it 6}, 15.

\bibitem{blochl}
Blochl,~P.~E. {\it Phys.~Rev.~B}, {\bf 1994}, {\it 50}, 17954.
 
\bibitem{paw}
Kresse,~G.; Joubert,~D. {\it Phys.~Rev.~B} {\bf 1999}, {\it 59}, 1758.
 
\bibitem{ldau2}
Bengone,~O.; Alouani,~M.; Bl\"{o}chl,~P.; Hugel,~J. {\it Phys.~Rev.~B},
{\bf 2000}, {\it 62}, 16392.
 
\bibitem{seq}
Schultz,~P.~A. http://dft.sandia.gov/Quest.

\bibitem{ultra}
Vanderbilt,~D. {\it Phys.~Rev.~B} {\bf 1990}, {\it 41}, 7892.
 
\bibitem{makov}
Neugebauer,~J.; Scheffler,~M. {\it Phys.~Rev.~B} {\bf 1992}, {\it 46}, 16067.

\bibitem{bengt}
Bengtsson,~L.  {\it Phys.~Rev.~B} {\bf 1999}, {\it 59}, 12301.

\bibitem{schul}
Schultz,~P.~A.  {\it Phys.~Rev.~B} {\bf 2000}, {\it 60}, 1551.
 
\bibitem{fcc}
The PBE lattice constant for FCC gold is 4.182\AA, in good
agreement with published lattice constants\cite{au_gga} computed
using the PW91 exchange correlation functional.\cite{pw91}
PW91 is very similar to PBE used in this work; both overestimate
the experimental lattice constant by 0.1~\AA.

\bibitem{au_gga}
Crljen,~Z.; Lazi\'{c},~P.; Sokcevic,~D.; Brako,~R. {\it Phys.~Rev.~B}
{\bf 2003}, {\it 68}, 195411.

\bibitem{mp}
Monkhorst,~H.~J.; Pack,~J.~D.  {\it Phys.~Rev.~B} {\bf 1976}, {\it 13}, 5188.
 
\bibitem{jaguar}
Jaguar v5.0, Schrodinger, LLC (Portland, Oregon, 2002).

\bibitem{ecp}
Russo,~T.~V.; Martin,~R.~L.; Hay,~P.~J. {\it J.~Phys.~Chem.} {\bf 1995},
{\it 99}, 17085.
 
\bibitem{g03}
Frisch,~M.~J. {\it et al.}, Gaussian 03 (Revision C.02), Gaussian Inc.,
(Wallingford, CT, 2004).

\bibitem{ahlrichs}
Schaefer,~A.; Horn,~H.; Ahlrichs,~R.  {\it J. Chem. Phys.} {\bf 1992},
{\it 97}, 2571.

\bibitem{wachters}
Wachters,~J.~H.  {\it J. Chem. Phys.} {\bf 1970}, {\it 52}, 1033;
Hay,~P.~J. {\it ibid}, {\bf 1977}, {\it 66}, 4377.
 
\bibitem{lammps}
Plimpton,~S. {\it J. Comp. Phys.} {\bf 1995}, {\it 117}, 1.

\bibitem{shel_para}
Song,~X.-Z.; Jaquinod,~L.; Jentzen,~W.; Nurco,~D.~J.; Jia,~S.-L.; Khoury,~R.~G.;
Ma,~J.-G.; Medforth,~C.~J.; Smith,~K.~M.; Shelnutt,~J.~A.  {\it Inorg.~Chem.}
{\bf 1998}, {\it 37}, 2009; Shelnutt~J.~A.; Medforth~C.~J.; Berber~M.~D.;
Barkigia,~K.~M.; Smith,~K.~M. {\it J.~Am.~Chem.~Soc.} {\bf 1991}, {\it 113},
4077.
 
\bibitem{rappe}
Rappe,~A.~K.; Casewit,~C.~J.; Colwell,~K.~S.; Goddard~III,~W.~A.; Skiff,~W.~M.
{\it J.~Am.~Chem.~Soc.} {\bf 1992}, {\it 114}, 10024.

\bibitem{note2}
Kozlowski {\it et al}.\cite{kozlowski} pointed out that, even
when using the {\it same} functional, localized basis Gaussian and
plane wave based calculations can yield discrepancies in the
low/intermedate spin splittings.  This may be a pseudopotential effect.
But the two methods are in agreement on the intermediate/high
spin splitting, which is what we use to fit DFT+U parameters.
Using plane wave basis VASP calculations, we have reproduced
Kozlowski {\it et al}.'s local basis PW91 predicted splittings
in Fe(II)P to within 0.04~eV.

\bibitem{kozlowski}
Kozlowski,~P.~M.; Spiro,~T.~G.; B\'{e}rces,~A.; Zgierski,~M.~Z.
{\it J.~Phys.~Chem.~B} {\bf 1998}, {\it 102}, 2603.

\bibitem{spin_contam}
Furthermore, the gas phase B3LYP results for intermediate and low spin states
show negligible spin contamination, and the DFT+U geometries for these
are similar to B3LYP predictions.

\bibitem{mn_expt2}
Boucher,~L.~J. {\it Coord.~Chem.~Rev.} {\bf 1972}, {\it 7}, 289, and
references therein.

\bibitem{mn_expt3}
Cheng,~B.; Scheidt,~W.~R.  {\it Acta Crystallogr. C.} {\bf 1996}, {\it 52},
361.
                                                                                
\bibitem{pd_expt1}
Fleischer,~E.~B.; Miller,~C.~K.; Webb,~L.~E. {\it  J.~Am.~Chem.~Soc.}
{\bf 1964}, {\it 86}, 2342.

\bibitem{pd_expt2}
Ishii,~T.; Aizawa,~N.; Yamashita,~M.; Matsuzaka,~H.; Kodama,~T.; Kikuchi,~K.;
Ikemoto,~I.; Iwasa,~Y.  {\it J. Chem.~Soc., Dalton Trans.}, {\bf 2000},
{\it 23}, 4407.

\bibitem{pd_expt3}
Seo,~J.-C.; Chung,~Y.-B.; Kim,~D. {\it Appl. Speectroscopy} {\bf 1987},
{\it 41}, 1199, and references therein.

\bibitem{pd_expt4}
Singh,~A.; Johnson,~L.~W.  {\it Spectrochimica Acta A}, {\bf 2003}, {\it 59},
905, and references therein.

\bibitem{ncd}
Jentzen,~W.; Song,~X.-Z.; Shelnutt,~J.~A.  {\it J. Phys. Chem. B} {\bf 1997}, 
{\it 101}, 1684.

\bibitem{ncd2}
For a recent review, see Shelnutt J.~A. in {\it The Porphyrin Handbook},
Kadish,~K.~M.; Smith,~K.~M.; Guilard,~R. eds.  (Academic Press, San Diego, 2000)
vol.~7, pp.~167.

\bibitem{ncd3}
Haddad,~R.~E.; Gazeau,~S.; P\'{e}caut,~J.; Marchon,~J.-C.; Medforth,~C.~J.;
Shelnutt,~J.~A. {\it J.~Am.~Chem.~Soc.} {\bf 2003}, {\it 205}, 1253.

\bibitem{feib}
Feibelman,~P.~J. {\it Phys.~Rev.~B} {\bf 2001}, {\it 64}, 125403,
and references therein.

\end{references}
\end{document}